\documentclass[12pt,preprint]{aastex}
\def\degpoint{\ifmmode ^{\rm{o}}\!. \else $^{\rm{o}}\!.$\fi}

\newcommand{\hd}{HD~209458b}
\newcommand{\kms}{\mbox{km \ s$^{-1}$}}
\newcommand{\Msun}{\mbox{M$_{\odot}$}}
\newcommand{\Rsun}{\mbox{R$_{\odot}$}}
\newcommand{\Mjup}{\mbox{M$_{\rm Jup}$}}
\newcommand{\Rjup}{\mbox{R$_{\rm Jup}$}}

\newcommand{\hst}{\textit{Hubble Space Telescope} }

\begin{document}
\title{System Parameters of the Transiting Extrasolar Planet
HD~209458b}
\author{Robert A.~Wittenmyer\altaffilmark{1}, William F.~Welsh, Jerome 
A.~Orosz}
\affil{Department of Astronomy, San Diego State University, San Diego, CA
92182}
\email{
robw@astro.as.utexas.edu,
wfw@sciences.sdsu.edu,
orosz@sciences.sdsu.edu}
\altaffiltext{1}{Department of Astronomy, University of Texas at Austin,
1 University Station, C1400, Austin, TX 78712}
\author{A.~B.~Schultz, W.~Kinzel, M.~Kochte}
\affil{Science Programs, Computer Sciences Corporation, and Space
Telescope Science Institute, 3700 San Martin Drive, Baltimore, MD 21218}
\author{F.~Bruhweiler}
\affil{IACS/Department of Physics, Catholic University of America, and
NASA Goddard Spaceflight Center, Greenbelt, MD 20771}
\author{D.~Bennum}
\affil{Physics Department, University of Nevada-Reno, Reno, NV 89557}
\author{Gregory W.~Henry}
\affil{Center of Excellence in Information Systems, Tennessee State
University, 330 10th Avenue North, Nashville, TN  37203; also Senior
Research Associate, Department of Physics and Astronomy, Vanderbilt
University, Nashville, TN 37235}
\author{G.~W.~Marcy, D.~A.~Fischer}
\affil{Astronomy Department, University of California at Berkeley, 601
Campbell Hall, Berkeley, CA 94720-3411}
\author{R.~P.~Butler}
\affil{Department of Terrestrial Magnetism, Carnegie Institution of
Washington, 5241 Broad Branch Road NW, Washington, DC 20015-1305}
\and
\author{S.~S.~Vogt}
\affil{University of California/Lick Observatory, University of California
at Santa Cruz, Santa Cruz, CA 95064}

\shorttitle{System Parameters of HD~209458b}
\shortauthors{Wittenmyer et al.}
\begin{abstract}

We derive improved system parameters for the HD~209458 system using a
model that simultaneously fits both photometric transit and radial
velocity observations. The photometry consists of previous 
\textit{Hubble Space Telescope} STIS and FGS observations, twelve 
$I$--band transits observed between 2001--2003 with the Mt.~Laguna 
Observatory 1m telescope, and six Str\"omgren $b+y$ transits observed 
between 2001--2004 with two of the Automatic Photometric Telescopes at 
Fairborn Observatory. The radial velocities were derived from 
Keck/HIRES observations. The model properly treats the orbital 
dynamics of the system, and thus yields robust and physically 
self-consistent solutions.
Our set of system parameters agrees with previously published results 
though with improved accuracy. For example, 
applying 
robust limits on the stellar mass of 0.93--1.20~\Msun, we find 
$1.26 < R_{\rm planet} < 1.42$~\Rjup\ and
$0.59 < M_{\rm planet} < 0.70$~\Mjup. 
We can reduce the uncertainty on these estimates by 
including a stellar mass--radius relation constraint, yielding
$R_{\rm planet}$=1.35$\pm$0.07~\Rjup\ and
$M_{\rm planet}$=0.66$\pm$0.04~\Mjup. 
Our results verify that the planetary radius is 10--20\% larger than 
predicted by irradiated planet evolution models, confirming the need 
for an additional mechanism to slow the evolutionary 
contraction of the planet. 
A revised ephemeris is derived,
$T_{0}=2452854.82545 + 3.52474554E$ (HJD), 
which now contains an uncertainty in the period of 0.016~s and should 
facilitate future searches for planetary satellites and other bodies 
in the HD~209458 system.

\end{abstract}

\keywords{stars: individual, HD~209458---extrasolar planets}
\section{Introduction}

Of more than 150 recently discovered extrasolar 
planets\footnote{A current tally can be found at the California and
Carnegie Planet Search homepage, http://exoplanets.org, or at the
J.~Schneider Extrasolar Planet Encyclopaedia,
http://www.obspm.fr/encycl/encycl.html.}, 
HD~209458b is the first known to transit its star \citep{henry00,
charb00}.  Combined with the spectroscopic radial-velocity curve,
photometric observations of transits allow high-precision determination
of system parameters such as the inclination, planetary radius and
mass.  For example, using \textit{Hubble Space Telescope} photometry of
HD~209458b, \citet{brown01} derived estimates of the inclination
$i=86\degpoint6 \pm 0\degpoint14$ and planetary radius
$R_{\rm planet}$=1.347 $\pm$0.060~\Rjup; 
with the additional estimate that the host star's mass is 
$M_{\rm star}$=1.06 $\pm$0.13~\Msun,
\citet{cody02} find $M_{\rm planet}$=0.69 $\pm$0.02~\Mjup.

With a measured radius and mass, HD~209458b observations are extremely
important in refining the theory of irradiated extrasolar giant planets
(EGPs). Most notably, the observed radius of HD~209458b is $\sim$10-20\%
larger than models predict \citep{gs02, bodenheimer01, bodenheimer03,
baraffe03, laughlin05}.  \citet{gs02} use evolutionary models of 
irradiated EGPs to argue for an additional heat source acting on 
HD~209458b.  They found that the radius of \hd\ can be produced with 
the transport of $\sim$1\% of incident stellar flux into the lower 
atmosphere as kinetic energy (i.e.~winds).  As \hd\ is probably tidally 
locked, one side is perpetually illuminated, and high-speed winds would 
be expected to transfer heat from the day side to the night side 
\citep{sg02, cho03, menou03}. \citet{bodenheimer01} suggest that 
continuing tidal circularization of \hd\ is heating the planet, 
inflating it to a radius larger than predicted for a circular orbit. 
However, a non--circular orbit is strongly disfavored by recent 
timings of the secondary eclipse of HD~209458b by \citet{deming05}. 
Observations of the transiting exoplanet TrES-1 
\citep{alonso04, sozzetti04, laughlin05}, indicating that its radius is 
consistent with models that do not invoke additional heat sources, 
support the suggestion of \citet{burrows04} that the large 
radius of HD~209458b 
may be anomalous. The radii estimates of the OGLE--discovered exoplanets 
are also consistent with models that include only irradiation.  
Determining the planetary radius of \hd\ more accurately would assist 
in resolving this discrepancy.

An extremely precise ephemeris can allow the inference of additional
bodies in the system by their gravitational effects on the transit times
and slight asymmetries in the transit light curve \citep{sartoretti99,
agol05, holman04}. The transit of \hd\ affords the opportunity to make
highly accurate and precise measurements of the planet's orbital period.  
The ability to predict and observe HD~209458b transits at 1-second
precision would place strong constraints on the masses of satellites or
additional planets in the system.  \citet{brown01} point out that an
Earth-mass satellite of HD~209458b would alter the time of mid-transit by
up to 13 seconds.  An ephemeris precise at the 1-second level would
facilitate tests for such timing displacements.  The precision of the STIS
transit timings obtained by \citet{brown01} allowed the exclusion of
satellites larger than 3 Earth masses at the 3$\sigma$ level by analysis
of the timing displacements of the four transits they observed.  Even if
no moons are present around HD~209458b, the technique of transit timing
can be readily applied to future transiting systems.  Some extrasolar
giant planets reside in the habitable zones of their parent stars, a
region where the star's insolation is such that liquid water could exist
on the surface of a planet.  Terrestrial-size moons of giant planets in
such an orbit could in theory be habitable \citep{williams97}.  While the
mid--transit timing perturbation caused by a hypothetical satellite 
orbiting HD~209458b
is on the order of 10~s, the perturbation grows as the cube root of the
planet's orbital period, so for a planet in the habitable zone the
perturbation due to a satellite could be tens of seconds to minutes in
duration.  The only currently available methods of detecting such 
satellites
require transits; refining modelling techniques using \hd\ thus builds a
solid foundation for characterization of future discoveries of transiting
extrasolar planets, e.g.~the \textit{Kepler} mission.  The most widely
used ephemeris by \citet{robichon00}, based on \textit{Hipparcos}
photometric data and the radial-velocity curve \citep{mazeh00}, contained
uncertainties in the period of $\pm1.21$ seconds. Propagating to the
present epoch yields an uncertainty in the time of mid-transit of over
$\pm$16 minutes.  There exists, therefore, a need to revise the ephemeris,
as the predicted time of mid-transit continuously accumulates errors in
period.  Even with a period determination accurate to 1 second, the large
number of cycles ($\sim$100 per year) results in nearly 2 minutes of
uncertainty accumulating after only a year.

In this work, we fit a single, self-consistent model to observations of
twenty-seven HD~209458b transit events in four bandpasses, and to more than
three years of high-precision radial velocities.  In \S 2 we briefly
describe the five data sets used in this study, \S 3 outlines the
modelling procedure, and in \S 4 we present and discuss the system
parameters.


\section{Observations}
\subsection{Mount Laguna Observatory (MLO)}

Twelve $I$--band transits of HD~209458b were observed with the 1-meter
telescope at Mount Laguna Observatory (MLO). A log of these and all other
photometric data is presented in Table 1.  Observing runs in 2002 and 2003
consisted of a 4 to 6 hour series of 1-second exposures on a Loral
$2048\times 2048$ CCD.  A field of view of 5x6.5 arcmin was used, which
included three faint comparison stars. Observations in 2001 used 4-second
exposures and a field of view of $5.9\times 5.9$ arcmin that only included
TYC 1688-1903-1 as a comparison star, and so this star was used as the
comparison on all nights.

The brightness of HD~209458 ($V$=7.645) required defocusing the telescope
slightly to avoid saturating the CCD; as such, large photometry apertures
($\sim 3 \times FWHM$ or 9 arcsec) were employed.  Light curves were
obtained using standard differential photometry with one comparison star.  
The photometry was then binned by a factor of 5--10 to a median time 
resolution of 177~s to help reduce the scintillation noise of the short 
exposures,
yielding a total of 1149 points. The error bars on the binned points were
calculated in the following way: uncertainties based on propagation of the
IRAF-generated error estimates were used to compute the inverse-variance
weighted mean per bin. Then this error bar was boosted by
$\sqrt{(\chi^{2}_{\nu})}$ if the reduced $\chi^{2}_{\nu}$ of the scatter
of the points about the mean in the bin was greater than 1.

For most transits, slight tilts ($\sim$0.5\%) in the light curves were
evident, likely due to systematic calibration errors.  Comparison star
variability was ruled out, as the tilts did not occur in all nights.  
Color-dependent differential extinction was also excluded, as the
correlation between the tilts and the airmass was not consistent.  These
tilts were corrected by masking the transit and fitting a line to the
out-of-transit light curve. This linear fit was then subtracted from the
data, rectifying the light curves. Finally, under the assumption that the
out-of-transit light curve is constant, the root mean square (RMS)
deviation of the out-of-transit observations was compared with the mean
uncertainty of these data. We found the error bars to be slightly
underestimated, almost certainly due to systematic errors unaccounted for
in the data reduction. We boosted all the uncertainties by 11\% to make
the uncertainties consistent with the out-of-transit RMS deviation.

\subsection{\hst STIS}

\citet{brown01} obtained and analyzed extremely high-precision
observations of four HD~209458b transits using the \hst Space Telescope
Imaging Spectrograph (STIS).  A total of 20 \hst orbits in four visits
spanning 18 days yielded a total of 684 spectra, covering a wavelength
range of $\sim$582--638 nm.  Binned over wavelength, the spectra yield
photometry with a relative precision of about $1.1\times10^{-4}$ per 60~s
integration. Following \citet{brown01} we omitted the first orbit of each
of the four HST visits in our modelling because of a systematic 0.25\%
deficit in flux, leaving 556 points in the light curve. As with the MLO
observations, the RMS deviation of the out-of-transit light curve about a
constant was used to check for the accuracy of the error estimates. We
boosted the uncertainties by 4.5\% to make the error estimates agree with
the RMS deviation out of transit. 

\subsection{\hst FGS}

Five transits were observed by \citet{schultz03} using photomultiplier
tubes (PMTs) 
in the \hst Fine Guidance Sensor (FGS).  The FGS, normally used for
astrometry and pointing control, was for the first time used as a
high-speed photometer on a bright star. The four PMTs in
FGS1r were used with a sampling rate of 40~Hz, and a S/N$\sim$80 per
0.025~s exposure was achieved. The F550W filter was used, giving a
central wavelength of approximately 550 nm. A time dependency in the FGS
response was corrected by fitting a 5th order Chebyshev polynomial to the
out-of-transit data for each transit and each PMT.  The data were then
placed into 80~s bins for consistency with the STIS data of
\citet{brown01}. 
Schultz et al.~(2004) have used these FGS data and the STIS data of
\citet{brown01} to obtain a precise set of
system parameters: $R_{\rm planet}$=1.367$\pm$0.043 \Rjup,
$R_{\rm star}$=1.154$\pm$0.036 \Rsun, and inclination
$i$=86\degpoint525$ \pm$0\degpoint054, assuming a stellar mass
of $M_{\rm star}=1.1\pm 0.1\,M_{\odot}$ from \citet{mazeh00}.
An orbital ephemeris of $T_0$=2452223.895819$\pm$0.000031 HJD and
$P$=3.52474408$\pm$0.00000029 days was also obtained; the uncertainty in
the period is only 0.025~s.  The precision of these data allowed the
exclusion of transiting satellites of HD~209458b down to 2 Earth radii. 
Such an object would cause a $\sim$0.1\% dip in the transit light curve,
which was not seen in the FGS data.  We trimmed a total of 19 data points
from the FGS observations, as these points were $>4\sigma$ outliers from
an initial model fit to the FGS data set.  These points occurred at the
beginning and end of each orbit, where the reliability of the FGS data is
the poorest, probably due to HST ``breathing.'' A total of 268 FGS
points were used in the modelling.  The uncertainty estimates of the
original Schultz et al.~(2004) data were boosted by 38\% to account for
systematic errors and be consistent with the out-of-transit RMS deviation
about a constant flux.

\subsection{Automatic Photometric Telescope (APT)}

Six transits were obtained in 2001--2004 with the T8 and T10 Automatic
Photometric Telescopes (APTs) located at Fairborn Observatory in the
Patagonia Mountains of southern Arizona (Henry 1999; Eaton, Henry, \&
Fekel 2003).  The transits of HD~209458b were co-discovered in 2000
using data from the T8 APT (Henry et al.~2000).  Precision photometers
use dichroic mirrors to split the incoming light into two beams and
two EMI 9124QB bi--alkali photomultiplier tubes to measure Str\"omgren 
$b$ and $y$ simultaneously.  Differential magnitudes from the two 
passbands were combined into a single $(b+y)/2$ band for greater 
precision, which is typically 0.0012~mag for a single measurement. 
The comparison star was HD 210074 ($V$=5.74 mag, F2V).  These APT data 
had a time resolution of 95~s and provide a total of 1426 observations.
The uncertainties were boosted by 7\% so as to be
consistent with the out-of-transit RMS deviation about a constant.

\subsection{Keck/HIRES Radial Velocities}

A set of 51 published \citep{henry00} and unpublished Keck/{\sc HIRES}
radial velocity measurements was included in the model fits and is listed
as Table 5 in the Appendix.  These data were obtained with the HIRES
echelle spectrometer at a resolution of $R \sim 80000$.  An iodine
absorption cell was used for wavelength calibration \citep{marcy92,
valenti95, butler96}. The current data set includes 42 measurements not
included in \citet{henry00}, and covers a time span of 3.4 years.  The
mean uncertainty of these velocity measurements is 4.7 m s$^{-1}$.

\section{Modelling of Transit Photometry and Spectroscopy}

\subsection{Outline of the Physical Model}

The Eclipsing Light Curve (ELC) code \citep{orosz00} was used to model
HD~209458 by simultaneously fitting all transit light curves and the 
radial velocities. ELC explicitly includes the orbital and rotational 
dynamics of the system, yielding robust and physically self-consistent 
solutions.
In ELC, the surfaces of the star and planet are defined by 
equipotential surfaces in the rotating binary frame. By using Roche 
equipotential surfaces to determine the radii, ELC can account for any 
non--sphericity of the bodies. In practice, for HD~209458 the
difference between the polar and equatorial radii is only 
$\sim$0.007\%, and can be neglected. ELC includes the effect of gravity 
darkening, though again because of the near sphericity of the star, 
this effect results in a negligible ($\sim$0.002\%) change in 
temperature between stellar pole and equator.

In the original version of ELC, two parameters called the 
``filling factors'' ($f_{\rm star}$ and $f_{\rm planet}$) were 
used to define the equipotential surfaces. The filling factors
are defined as the ratios of radii to Roche lobe radii, such
that $f<1$ for detached systems and $f=1$ for a Roche lobe--filling 
body. For given masses, the filling factors thus define the stellar 
and planetary radii. Unfortunately, this way of specifying the
equipotential surfaces is far from optimal in the case of HD~209458. 
Since the two bodies are nearly perfect spheres, a change in the 
mass ratio $Q=M_{\rm star}/M_{\rm planet}$ requires a change in 
$f_{\rm star}$ \textit{and} $f_{\rm planet}$ to get the same transit 
profile.  
Thus it proved to be much more computationally efficient to
have as the two parameters the {\em fractional radius}
of the star $R_{\rm star}/a$ and the {\em ratio of the radii}
$R_{\rm star}/R_{\rm planet}$ since the transit light curve is
independent of the mass ratio when the equipotential surfaces 
are defined this way.

Once the surfaces of the star and planet are defined they are then
divided into a grid of surface elements (``tiles''). 
For HD~209458 a very fine grid of surface elements was required: 
even with  $400\, {\rm latitude} \times 600\, {\rm longitude}$
tiles on the star we found numerical noise in the light curves of the
order of 1 part in $10^{5}$, which is not good enough to model the 
precision of the \hst light curves.
To mitigate this effect, we employed a 
Monte Carlo subsampling of the partially eclipsed tiles, 
using 1000 random subsamples per tile.
This technique allowed a much smaller number of grid
elements and we adopted a final $250\times 240$ 
grid on the star and $80\times 80$ grid on the planet.

The intensity at each tile is approximated as a blackbody, accounting for
limb darkening and correcting the effective temperature $T_{\rm eff}$ for
gravity darkening \citep{vonzeipel24, claret00}.  Limb darkening was
treated using the 2-parameter logarithmic prescription of 
\citet{vanhamme93}:
\begin{equation}
I(\mu)=I_0(1-x+x\mu -y\mu \textrm{ln} \mu)
\end{equation}
where $\mu$ is the foreshortening angle of the grid element and
$x$ and $y$ are the two limb-darkening coefficients for each bandpass.
Blackbody intensities were employed for convenience, but also because the
Kurucz model atmosphere tables are presently too coarse at the
$T_{\rm eff}$=6000~K temperature appropriate for HD~209458 
(\citet{mazeh00}, \citet{cody02}).
After correcting the intensity of each tile for limb and gravity
darkening, the binary is then ``turned in space'' by a user-specified
phase step. A phase step size of 0\degpoint05 was chosen,
corresponding to about 42.3~s, to match the high time resolution
(80~s) and precision
($\sim1.1\times10^{-4}$) of the \hst data.
If a tile on the star is completely eclipsed by the planet, that 
hidden 
tile is not included in the
calculation of total flux from the star.  Partially eclipsed tiles are
accounted for via the Monte Carlo method mentioned above. In this way, a
model light curve is generated for the input parameters.  

In the original version of ELC, the scale of the binary
was fully specified by using the inclination $i$, the
orbital separation $a$, and the mass ratio $Q$.  In the case of
HD~209458, we found it more convenient to use  the mass
of the star $M_{\rm star}$ and the projected  semiamplitude of
the star's radial velocity curve $K_{\rm star}$.  For a given
inclination $i$, the separation $a$ and the mass ratio 
can be determined from $M_{\rm star}$ and $K_{\rm star}$.  
Then given the orbital period $P$, 
the radial velocity of the star at each orbital phase is computed by 
summing the velocities of each stellar tile not eclipsed by the planet, 
again weighting for limb darkening and foreshortening \citep{wilsof76}. 
In this way we fit the radial velocity observations, 
and do not just constrain the stellar motion to equal some adopted
projected orbital velocity $K_{\rm star}$. 

Since ELC computes the projected radial velocities at each orbital 
phase, the Rossiter effect can be fit with our model. The Rossiter 
effect is a radial--velocity distortion which occurs as a result of the 
planet blocking the approaching and receding limbs of the star during 
transit (Rossiter 1924).  The Rossiter effect was first observed in 
HD~209458 by \citet{bundy00} and \citet{queloz00}, and was used to 
deduce that the planet orbits in the same direction as the star's 
rotation.  The amplitude of the Rossiter effect is directly 
proportional to the star's rotational velocity $V_{\rm rot} \sin{i}$ and 
the size of the eclipsing object.  As portions of the rotating stellar 
surface are blocked by the planet, the absorption line profiles are 
skewed in shape.  Thus, the velocities reported here during the transit 
represent an apparent Doppler shift caused by the \textit{skewing} of 
the original absorption lines, rather than a change in velocity. 
Because of this, we caution that the radial velocities through the 
transit as computed by ELC may not be identical to the radial velocities
as measured in the stellar spectra via the cross--correlation method. 
A thorough investigation would require the convolution of a high 
signal--to--noise out--of--transit spectrum (or model spectrum) with a 
set of phase--dependent line--broadening functions generated during the 
transit. These synthetic spectra should then be cross--correlated with 
a template spectrum and the resulting simulated radial velocities 
compared with the ELC radial velocities. Such an investigation is 
beyond the scope of this paper, but we note that the Rossiter effect 
observations are in fact very well matched by the ELC model as seen in
\S 4.2.

A circular ($e=0$) planetary orbit was assumed, consistent with a tidal
circularization time of order $10^{8}$ yr for HD~209458b
\citep{bodenheimer03} and radial velocity observations \citep{mazeh00}.  
Recent {\em Spitzer Space Telescope} observations by \citet{deming05}
show the secondary eclipse occurs at orbital phase
$0.5000\pm 0.0015$, demanding a very low eccentricity.
The planet is also assumed to be tidally locked (rotation synchronous 
with its orbit) so the rotational frequency parameter $\Omega_{\rm 
planet}$, defined as the ratio of rotational frequency to the orbital 
frequency, was set to unity.  The effective temperature of the planet 
was set to 1300~K, based on theoretical estimates of the night-side 
temperature of irradiated extrasolar giant planets (\citet{gs02}; 
\citet{deming05} measure a brightness temperature of $1130\pm 150$ K). 
At this temperature the flux ratio of the planet to star is 
approximately $3 \times 10^{-11}$ in the $V$ band, and so the planet is 
essentially invisible in our optical photometry.

Finally, note that ELC computes the actual orbital path of the planet 
about the barycenter rather than simply a chord across the stellar 
disk. This includes the reflex motion of the star during transit, which 
for a 3.07~hr transit, amounts to a distance equal to 0.9\% the radius 
of the planet. Though these effects are subtle, if improperly treated 
they can be sources of systematic error and can bias the system 
parameters. By including the dynamics of the orbits, ELC should in 
principle yield more accurate, as well as precise, system parameters.

\subsection{Observational Constraints}

HD 209458 is a single-lined binary system, and it is well-known that 
the radial velocity curve of the star sets the minimum mass of the 
planet:
$$
f(M_{\rm planet})\equiv {PK_{\rm star}^3\over 2\pi G}=
{M_{\rm planet}^3\sin^3i\over
(M_{\rm planet}+M_{\rm star})^2}.
$$
Once the mass function $f(M_{\rm planet})$ is known, one needs the 
inclination and the mass of the star (or the ratio of masses) in order 
to find the mass of the planet $M_{\rm planet}$.  Once the component 
masses are known, the orbital separation $a$ can be found from Kepler's 
third law.
Since the transit light curves constrain the ratio of the radii 
($R_{\rm star}/R_{\rm planet}$) and the fractional radii 
($R_{\rm star}/a$ and $R_{\rm planet}/a$), the radii of the star and 
planet in physical units are found once $a$ is known.

From a strictly observational point of view, the inclination, relative
radii, and limb darkening are constrained quite well from the 
multi-color transit light curves. However, neither the stellar mass or 
the mass ratio is constrained by the transit or radial velocity 
observations and so the scale of the system is indeterminate (see also 
the discussion in \citet{deeg01}).
Since we are interested in the physical size and mass of the planet, an 
additional constraint is required to break the degeneracy.
Fortunately we do have constraints on the radius and mass of the host 
star, by matching stellar evolution models with the star's 
luminosity and temperature.

\citet{cody02} derive a stellar radius of 1.18$\pm$0.10 \Rsun\ using 
the apparent $V$ magnitude, the {\em Hipparcos} parallax
\citep{perryman97}, and a well-determined bolometric correction. 
The $\sim8\%$ error on this stellar radius estimate yields an 8\% 
error on the orbital separation and hence will yield an uncertainty of 
$\pm$8\% in the planetary radius.
We include this stellar radius estimate as an observation to be matched 
by the ELC model. The model is directed toward this value by use of a 
straightforward  $\chi^{2}$ penalty for deviations from the estimate. 
In a similar fashion, we include the observation of the projected stellar 
rotation velocity $V_{\rm rot} \sin{i}$ of \citet{queloz00} as a datum 
to be matched. Thus in addition to the transit photometry and radial 
velocities, we include two more observables, with the radius of 
HD~209458 ``steered'' toward 1.18$\pm$0.10 \Rsun, and toward a 
$V_{\rm rot} \sin{i}$ of  3.75$\pm$1.25 \kms. However, we emphasize 
that these are not constraints --- the model is permitted to adopt 
values outside of these ranges, but at a cost in $\chi^{2}$.

\citet{cody02} also determined a robust limit on the stellar mass of 
1.06$\pm0.13$ \Msun; this uncertainty range includes the observational 
errors in temperature, luminosity, and metallicity as well as systematic 
errors in convection mixing--length and helium abundance. 
The $\pm$12\% uncertainty on the stellar mass translates to roughly a 
$\pm$4\% error on the planet radius. 
However, this range of allowable stellar mass is not quite a   
1-$\sigma$ uncertainty, but more of a limit on the range of possible 
mass. For this reason, and also because the stellar mass determination
is more model-dependent than the other parameters, we do not include
the stellar mass as an observed parameter.

The relative uncertainty in the orbital velocity $K_{\rm star}$ also maps
into an uncertainty in orbital separation and hence radii. The
correspondence is 1:1, so an uncertainty of 2.1\% (see \S 4.2)
in $K_{\rm star}$ gives a {\em minimum} uncertainty of 2.1\% in radii.
This 2.1\% is the uncertainty in the radii {\em if} the mass of the star
were exactly known.

\subsection{Fitting Procedure}

Using the above assumptions and conditions, all four photometric time
series were modelled simultaneously with the radial velocities in order to
obtain system parameters for HD~209458.  The criteria defining the
goodness of fit was the usual $\chi^{2}$. The ELC model contains 15 free
parameters:  the orbital period of the planet $P$, the time of mid-transit
$T_{0}$, the fractional radius of the star $R_{\rm star}/a$, the ratio of
the radii $R_{\rm star}/R_{\rm planet}$, the inclination, the
semiamplitude of the star's radial velocity curve $K_{\rm star}$, the
ratio of rotational frequency of the star to orbital frequency, and eight
wavelength-dependent limb-darkening coefficients (two per bandpass).
Because of the degeneracies in the solutions, the stellar mass was not a
free parameter in the models. Rather, we fix the stellar mass and optimize
all other parameters, then repeat with a different stellar mass. In this
way we ran 69 models, stepping through stellar masses between 0.57 and 
1.72 \Msun. The list of other fixed input parameters and their values is
given in Table~2.
 
The weighting of the five data sets was determined by their 
uncertainties; no adjustments to the relative weights were made.  
In order to assure complete exploration of parameter space, a genetic 
algorithm based on the {\em pikaia} routine given in 
\citet{charbonneau95} was used to find the global $\chi^2$ minimum (see 
\citet{orosz02} for details), and a  downhill simplex ``amoeba'' 
\citep{press} was used to further examine the minimum.
A simple grid search was then used to step through the stellar mass.

For the purposes of this investigation, the input light curves were 
scaled in the following way: APT data were treated as Johnson $B$ band 
light curves, FGS data were considered as $V$ band, STIS data were 
considered $R$ band, and MLO data were treated as $I$ band.  Using the 
\textit{Hipparcos V} magnitude of 7.645 \citep{perryman97} and the 
observed $V-I$, $R-I$, and $B-V$ color indices \citep{hog00}, the 
out-of-transit fluxes were scaled to $B=8.18$, $V=7.645$, $R=7.287$, 
and $I=6.985$ mag.  Although the effective bandpasses of the four light 
curves are not equivalent to Johnson filters, these approximations are 
acceptable since ELC compensates by adjusting the limb darkening 
parameters.  Also note that the bandpasses are treated independently, 
i.e.~the relative fluxes between them are not constrained by the 
blackbody temperatures, so the out-of-transit scaling can in fact
be arbitrary.


\section{Results and Discussion}

\subsection{Light Curve Fits}
Figure 1 shows plots of all photometric data phase-folded and overlaid
with the ELC model light curves.  For this figure, the APT and MLO light
curves were binned by a factor of 5 for clarity. 
The effect of color-dependent limb darkening is evident in the shape of
the transits: the I-band light curve is considerably flatter during
mid-transit than the B-band. 
The reduced $\chi^2$ of each data set
are the following: 
MLO:     1.96,
STIS\footnote{
If we fit to the full STIS light curve, including the out-of-transit data
obtained during the first HST orbit of the visit, the STIS reduced 
$\chi^2$ is 1.54; this is a consequence of the systematically lower
flux for these unreliable data, as described in \citet{brown01}. 
}:       0.95,
FGS:     1.27,
and APT: 1.01.
The reduced $\chi^2$ of the entire fit to all the observations
is 1.36 with 3434 degrees of freedom.
The reduced $\chi^2$ values for the STIS and APT observations are
excellent; for the MLO and FGS observations the $\chi^2$ is high,
probably as a consequence of systematic errors in the data rather than
deficiencies in the model.  Normalized residuals ($(O-C)/\sigma$) of the
model fits are shown in Fig.~2 and, with the exception of the MLO
data, show no pattern of correlated residuals. 
The MLO residuals do exhibit structure: just prior to mid-transit the data
are generally above the model, and after mid-transit the data are below
the model. We are confident this is an observational calibration problem,
perhaps a residual remaining after the linear rectification.

\subsection{Radial-Velocity Fit}

Shown in Figure 3, the model radial velocities match the observations
throughout the orbit. No obvious patterns indicative of a nonzero
eccentricity are evident in the residuals.  The reduced $\chi^2$ of this
model fit to the radial velocities is 2.16, though removal of one outlier
at phase 0.05 lowers the reduced $\chi^2$ to 1.80.  The RMS scatter of the
observations about the fit is 6.6 m s$^{-1}$, not much larger than the
mean uncertainty of 4.7 m s$^{-1}$ in the data. We derive a stellar 
reflex velocity $K_{\rm star}=82.7\pm 1.3$ m s$^{-1}$
(note that this is independent of the assumed stellar mass). 
This is in good agreement with the results of 
\citet{henry00} 81.5$\pm$5.5 m s$^{-1}$, 
and marginally smaller than the values of 
\citet{mazeh00} 85.9$\pm$2.0 m s$^{-1}$ and 
\citet{naef04} 85.1$\pm$1.0 m s$^{-1}$.

Figure 3 also shows a detailed plot of the fit to the 
Rossiter effect, showing the excellent agreement between the ELC model 
and the observations.
During the transit the RMS of the residuals is 3.9 m s$^{-1}$, 
whereas the out-of-transit RMS is 8.5 m s$^{-1}$.  The data
taken during transit were all obtained on the same night (JD 2451755),
whereas the remaining data come from widely separated epochs spanning 
more than three years.  Systematics such as the long-term stability of 
the spectrograph are thus most likely the cause of the increased scatter
outside of transit.

\subsection{The Orbital Ephemeris of HD~209458b}

While extremely precise, the STIS observations only span 6 nearly
consecutive HD~209458b orbital cycles and therefore are not ideal for
determining the ephemeris. The FGS observations span 1.3 years and hence
place much tighter constraints on the period.  By combining the STIS and
FGS high-precision data with the ground--based observations spanning many
cycles, very tight limits can be placed on the orbital period. We obtain a
revised ephemeris of $T_{0}=2452854.82545+3.52474554E$ HJD (Table 3). The
uncertainty in period is now only 0.016~s, compared to 1.21~s
\citep{robichon00}, an improvement by over a factor of 70.  A 
comparison of our ephemeris with previously published results is shown 
in the O-C  diagram of
Fig.~4, which displays the residuals of times of mid-transit from our
ephemeris.  We include most previously published mid-transit times
\citep{charb00, jha00, mazeh00, deeg01, schultz03, schultz04}.  
Since we have included the STIS and FGS data in our fits, the estimates of
$T_0$ by \citet{schultz04, schultz03} and \citet{brown01} are not 
independent and so we cannot use the $O-C$ values in Fig.~4 to further 
improve the ephemeris.

Our results are in general agreement with previous work, but some
discrepancies exist; for example, our period differs from
\citet{schultz04} by $\approx $8 $\sigma$ ($\sim 0.126$ sec). The
majority of the uncertainty in the ephemeris lies in $T_0$; the
uncertainty in $T_0$ is 750 times larger than that in the period.  
Clearly, if $T_0$ can be determined as accurately as $P$, attaining
1-second precision in transit timings, and hence searches for satellites
and other gravitationally perturbing bodies, would be far less
challenging. Despite their very high photometric precision, neither the
FGS or STIS can obtain an uninterrupted observation of the 185~min transit
due to HST's low Earth orbit. This makes determining the center of transit
much more difficult, and hence more uncertain. What is needed, then, are
observations of \textit{complete} transits at a precision comparable to
the \textit{Hubble Space Telescope} data, or a large set of
lower-precision mid-transit times spanning many cycles. The Canadian {\it
MOST} satellite, expected to achieve micromagnitude photometric precision,
is well-suited for this task, as its orbit enables it to continuously
observe HD~209458 for many transits \citep{rucinski03} with exceptionally
high precision.

\subsection{HD~209458 System Parameters}

Given only the transit light curves and radial velocities, the absolute 
physical scale of the system cannot be determined and the solutions are 
degenerate. 
In Fig.~5 we show the observed relationships between the planetary 
and stellar radii versus the stellar mass. The solid dark curves are 
the ELC solutions that best match the transit and radial velocity data, 
and as expected, the curve follows $R \propto M_{\rm{star}}^{1/3}$. 
Formally, a fit to the model points yields the relation 
$R_{\rm planet}= 1.310 \times (M_{\rm star})^{0.334}$ \Rjup\ and 
$R_{\rm star} =  1.112 \times (M_{\rm star})^{0.333}$ \Rsun, and these are 
shown as the dark curves in Fig.~5.
The planetary and stellar radii are observationally constrained to lie
along these curves, though their positions on the curve are only weakly 
constrained.
By including Cody \& Sasselov's (2002) observationally-derived stellar 
radius 
of 1.18$\pm$0.10 \Rsun\ we break the degeneracy, but the uncertainty in 
the system scale remains large, giving a minimum uncertainty of $\pm$8\% 
in the planetary radius and 25\% in the stellar mass. Our best fit value 
for the stellar mass using only the stellar radius constraint (along 
with the transit photometry, radial velocities and $V_{rot} \sin i$) 
is 1.25$\pm$0.35 \Msun.

If we apply the \citet{cody02} stellar mass limit constraint, 
1.06$\pm$0.13~\Msun, we limit the solutions to the light gray region
in Fig.~5. The smaller, darker gray region corresponds to the mass 
range given by \citet{santos04}, 1.15$\pm$0.05~\Msun. The vertical 
width of the gray regions correspond to the $\pm$2.1\% uncertainty in 
radius derived from the uncertainty in $K_{\rm star}$. With these mass 
range estimates, we can robustly bracket the planetary radius to the 
range 1.28 -- 1.42 \Rjup. 
Note that this is not a statistical 1--$\sigma$ confidence interval;
solutions outside this range are highly disfavored.
For comparison, we show the planetary radius estimates of 
\citet{cody02}, \citet{deeg01}, \citet{brown01}, and \citet{schultz04}. 
The latter three studies assumed a stellar mass of 1.10 \Msun\ to get the
planetary radius, and so their radius ranges should be compared to the 
height of the gray region on our radius vs.~mass curve. The box and 
diamond symbols show the locations of the points corresponding to the 
best stellar mass estimates of \citet{cody02} and \citet{santos04}.

The lower panel of Fig.~5 shows the family of stellar radius vs.~stellar 
mass solutions for HD~209458. This relationship comes solely from the 
observations of the transit and radial velocities, and does not include 
any stellar astrophysics. From the transit and radial velocity  
observations alone, there is no difference in goodness of fit 
anywhere on this curve.  The mass--radius relationship for HD~209458 
given by \citet{cody02}, based on stellar models, is shown as the 
dot--dash line (bracketed by the uncertainty in the stellar radius). 
As \citet{cody02} noted, this M--R relationship is nearly orthogonal to 
the transit curve, and thus introduces a strong constraint on the 
stellar mass and radius. The intersection of the \citet{cody02} M--R 
relationship with our transit--derived relationship occurs at 
1.09 \Msun\
and defines our best estimate of the stellar mass. 
Incorporating the uncertainty in the mass--radius relation then yields
these values: 
$M_{\rm star}  $=1.093 $\pm$0.092~\Msun,
$R_{\rm star}  $=1.145 $\pm$0.056~\Rsun, and
$R_{\rm planet}$=1.350 $\pm$0.066~\Rjup, respectively,
though again we caution that the uncertainties are not statistical 
1--$\sigma$ confidence intervals.

We estimate a slightly higher stellar mass and smaller stellar radius 
than \citet{cody02}. This requires a slightly different set of adopted 
system parameters, e.g., we find the inclination to be 86.67 versus 
86.1 degrees. But more importantly, our smaller stellar radius requires 
a higher temperature to produce the observed luminosity. Interestingly,
\citet{santos04} estimate the temperature to be 6117$\pm26$ K, 
significantly hotter than the \citet{cody02} or \citet{mazeh00} value 
of 6000$\pm$50 K. \citet{ribas03} also find a somewhat higher 
temperature, 6088$\pm$56 K, as do \citet{fischer-valenti05},
6099$\pm$44 K.
Even a small temperature increase of 88~K requires a 3\% change in 
radius to maintain the same bolometric luminosity, and would 
reduce the radius from the \citet{cody02} value of 1.18 \Rsun\ to 
1.145 \Rsun; a 117 K increase in effective temperature would drop the 
radius to 1.136 \Rsun; both of these agree well with our stellar radius 
estimate, suggesting stellar temperatures greater than 6000 K are 
favored. These higher temperatures imply a slightly earlier spectral 
type, F9, than the usual adopted value of G0
(see e.g.~\citet{gray-napier-wink} and \citet{gray-graham-hoyt}).

A higher temperature would also mean a younger age, as can be seen in the
stellar evolution tracks shown in Fig.~1 of \citet{cody02}: the
observation box would be shifted to the left, and closer to the main
sequence. However, the slightly younger age makes little difference to the
planetary radius predicted from planetary evolution, since the shrinking
of the planet vs.~time is quite slow after $10^9$ years
e.g.~see \citet{burrows01}.

As a check, we estimated the stellar radius from the {\em Hipparcos} 
parallax--derived distance, the observed magnitudes (assuming zero 
reddening), the observed surface gravity \citep{santos04}, an 
assumed effective temperature, and synthetic photometry computed from 
the {\sc NextGen} stellar models \citep{hauschildt99a}. 
These stellar radius estimates are shown in Fig.~5, 
for $T_{\rm eff}$=6000 K and 6117 K. 
Also shown are the stellar radius estimates of 
\citet{cody02}, \citet{allende-lambert}, and \citet{ribas03}; 
all are consistent within their uncertainties and with our ``best'' 
mass and radius estimate.
In particular, the optical--IR synthetic photometry technique employed 
by \citet{ribas03} gives a stellar radius of 1.145$\pm$0.049 \Rsun\, 
in excellent agreement with our favored estimate.

It is interesting that the \citet{cody02} planetary radius estimate is 
notably larger than our value, despite our heavy reliance on their 
stellar radius estimate to set the physical scale of the system.
Using our transit--derived radius vs.~mass relation, 
their stellar radius estimate (1.18 \Rsun) is inconsistent with their
preferred stellar mass (1.06 \Msun); a larger stellar mass,  
1.20~\Msun, is required for consistency with our transit modelling.
For completeness, we note that the radius vs.~mass relationship 
derived by \citet{deeg01},
$R_{\rm star}=0.34 M_{\rm star} + 0.825 \ (\pm0.06)$, 
gives a stellar radius larger by $\sim$0.05 \Rsun\ than our results, 
and hence their system scale is larger by approximately this amount. 
Their estimate was made prior to the exquisite \hst transit 
observations and the tighter constraints those observations 
provide; our stellar radius vs.~mass relation is similar to their 
lower 1--$\sigma$ limit on the relation.

Just as with the planetary radius, the planetary mass depends on the
adopted stellar mass. The transit and radial velocity data constrain 
the planetary mass to lie along a curve; formally we find
$M_{\rm planet}$=$0.620 \times (M_{\rm star})^{0.670}$ \Mjup\ from
our model fits. This is exactly what is expected for circular Keplerian 
orbits and mass ratio $Q \gg 1$:
$M_{\rm planet}$=$(P / 2 \pi G)^{1/3} (K_{\rm star}/ \sin{i}) \times 
(M_{\rm star})^{2/3}$.
Applying the stellar mass--radius constraint as above, 
we find $M_{\rm planet}$=0.658 $\pm$0.036~\Mjup.

In Table 4, we present the system parameters as a function of 
stellar mass for three representative cases,
$M_{\rm star}$ = 0.93, 1.09, and 1.19 \Msun. 
These cases bracket the full range in stellar mass that is deemed 
acceptable, following \citet{cody02}.
The overall reduced $\chi^2$ of the fit to the 3434 observations
(transit photometry, radial velocities, $R_{\rm star}$, and 
$V_{\rm rot} \sin{i}$) was 1.364. Because of the degeneracy of the 
solutions discussed above, the $\chi^2$ does not vary significantly 
across this mass range.

Our system parameter values are generally in good agreement with those
derived in previous work. Our estimate of the planetary radius is strongly
constrained to be between 1.26 -- 1.42~\Rjup, using the stellar mass
limits of \citet{cody02} and \citet{santos04}. Applying the stellar
mass--radius relation of \citet{cody02}, we get the limits:  1.35
$\pm$0.07~\Rjup. This radius is consistent with previously published
results \citep{brown01, cody02, schultz04, deeg01, laughlin05}, and hence
the discrepancy between the observed and theoretical models of extrasolar
giant planets remains. The planetary radius is 10--20\% larger than
evolutionary models which only include irradiation \citep{chabrier04};
these give a radius of $\sim$1.1~\Rjup, equal to the bottom edge of the
top panel in Fig.~5, and firmly ruled out.  However, Burrows et al.~(2003)
point out that the observed transit radius (radius where the planet's
optical depth to starlight transmitted through the atmosphere along our 
line of sight is unity) is not
the same as the theoretical 1 bar radius, and the difference can
account for as much as 0.1 \Rjup  in the radius. When this effect is
included, irradiated models can be marginally consistent with the extreme
lower limit on the observed radius.  As an alternative explanation for the
planet's large radius, \citet{bodenheimer01} suggested that eccentricity
pumping by an undetected additional planet could provide the energy needed
to slow the evolutionary contraction of HD~209458b. However, the
radial-velocity residuals currently do not support a significant nonzero
eccentricity.  The recent observation of the secondary eclipse of
HD~209458b by \citet{deming05} also strongly supports an eccentricity
indistinguishable from zero, as the timing of the secondary eclipse is not
significantly displaced from phase 0.5.  Thus the observed radius of
HD~209458b remains unsatisfactorally explained and continues to
demonstrate deficiencies in our understanding of irradiated extrasolar
giant planet evolution.  Finally, with a mass of 0.66 $\pm$0.04~\Mjup\ and
the above radius, the mean density of HD~209458b is
0.33$\pm$0.05~g~cm$^{-3}$, about half the density of Saturn and one-fourth
the density of Jupiter; HD~209458b is a very low density gas giant planet.

\section{Summary}

We have used the ELC code \citep{orosz00} to determine the system 
parameters of HD~209458 by simultaneously fitting the transit light 
curves and radial velocities. The observations consist of the \hst 
STIS \citep{brown01} and FGS \citep{schultz04} light curves, plus 
18 transits obtained over 4 years with the facilities at MLO and APT, 
along with Keck {\sc HIRES} spectroscopic radial velocities. Our new 
estimates of the system parameters are generally in agreement with 
previous results, e.g.,
1.26 $< R_{\rm planet} <$1.42 \Rjup\ and
0.59 $< M_{\rm planet} <$0.70 \Mjup.
We stress that this range includes the uncertainty in the stellar 
mass. By applying the mass--radius relation of \citet{cody02}, 
we reduce the uncertainty:
$R_{\rm planet}$=1.35 $\pm$0.07~\Rjup,
$M_{\rm planet}$=0.66 $\pm$0.04~\Mjup.
Our results confirm that the planetary radius remains significantly 
larger (10--20\%) than predicted by irradiated planet evolution models 
e.g.~see \citet{chabrier04}. 
For the stellar parameters, we find
$M_{\rm star}  $=1.09 $\pm$0.09~\Msun\ and
$R_{\rm star}  $=1.15 $\pm$0.06~\Rsun.

We have also obtained an orbital ephemeris with a period determination
good to 0.016~s, over 70 times more precise than the period by
\citet{robichon00}.  An ephemeris of this precision should facilitate
future searches for additional bodies in the HD~209458 system.  
By using ELC, a
fully self-consistent dynamical model that includes subtle physical
effects not contained in previous work, we have reduced systematic errors.
As observational precision increases, and detection methods become
sensitive to smaller planets at larger orbital distances (ultimately to
Earth analogs), the need for such exactitude in the modelling is
warranted.

\acknowledgements
We are grateful to an anonymous referee whose comments improved this 
manuscript.  The authors wish to thank Drs.~E.~L.~Robinson, K.~Horne and 
P.~Hauschildt for useful discussions related to this work.
We also thank Ms.~D.~Martino and S.~Airieau for assistance in 
obtaining transit observations at MLO.
GWH acknowledges support from NASA grant NCC5-511 and NSF grant HRD
97-06268, and WFW acknowledges support by an award from 
Research Corporation.
This research has made use of NASA's Astrophysics Data System (ADS), 
the SIMBAD database, operated at CDS, Strasbourg, France, NASA's 
SkyView (http://skyview.gsfc.nasa.gov) and Dr.~J.~Thorstensen's 
SkyCalc software.
Observations obtained at MLO made use of the HPWREN (``High 
Performance Wireless Research and Education Network'') sponsored by 
the NSF ANIR division under grant ANI-0087344 and the University of 
California San Diego. We are greatly indebted to D.~Charbonneau et 
al.~for providing the \hst STIS photometry.  The observations used in this 
work can be obtained from W.~Welsh at wfw@sciences.sdsu.edu.


\begin{deluxetable}{lll} 
\tabletypesize{\scriptsize}
\tablecolumns{3}
\tablewidth{0pt}
\tablecaption{Observation Log \label{tbl-1}}
\tablehead{
\colhead{Instrument} & \colhead{UT Date} & \colhead{Start Time
(HJD-2450000)}}

\startdata
MLO & 2001 Jun 29-30 & 2089.840 \\
MLO & 2001 Oct 13 & 2195.636 \\
MLO & 2002 Jun 13 & 2438.836 \\
MLO & 2002 Aug 5 & 2491.736 \\
MLO & 2002 Aug 12 & 2498.726 \\
MLO & 2002 Sep 27 & 2544.626 \\
MLO & 2002 Oct 4 & 2551.640 \\
MLO & 2002 Oct 11 & 2558.616 \\
MLO & 2002 Oct 18 & 2565.681 \\
MLO & 2002 Dec 10 & 2618.585 \\
MLO & 2003 Jul 27 & 2847.776 \\
MLO & 2003 Aug 3 & 2854.718 \\
STIS & 2000 Apr 25 & 1659.744 \\
STIS & 2000 Apr 28-29 & 1663.297 \\
STIS & 2000 May 5-6 & 1670.336 \\
STIS & 2000 May 12-13 & 1677.376 \\
FGS & 2001 Jun 11 & 2072.274 \\
FGS & 2001 Sep 11-12 & 2163.902 \\
FGS & 2001 Nov 10 & 2223.793 \\
FGS & 2002 Jan 16 & 2290.783 \\
FGS & 2002 Sep 30 & 2548.085 \\
T8 APT & 2001 Oct 6 & 2188.592 \\
T8 APT & 2001 Oct 13 & 2195.579 \\
T10 APT & 2002 Oct 4 & 2551.589 \\
T10 APT & 2002 Oct 11 & 2558.581 \\
T10 APT & 2004 Sep 15 & 3263.606 \\
T10 APT & 2004 Sep 22 & 3270.597 \\

\enddata
\end{deluxetable}

\begin{deluxetable}{lll}
\tabletypesize{\scriptsize} 
\tablecolumns{3}
\tablewidth{0pt}
\tablecaption{ELC Model Input Constraints \label{tbl-2}}
\tablehead{
\colhead{Parameter} & \colhead{Value} & \colhead{Reference or Reason}}

\startdata 
$T_{\rm eff}$, star & 6000 K & Cody \& Sasselov (2002)\\
$T_{\rm eff}$, planet & 1300 K & Guillot \& Showman (2002)\\
Orbital eccentricity $e$ & 0.00  & Mazeh et al.~(2000)\\
Planet rotation/orbital frequency ($\Omega_{\rm planet}$)
                          & 1.00  & Assume tidal locking\\
\hline
Radius of star $R_{\rm star}$ & 1.18$\pm$0.10 \Rsun & Cody \& Sasselov
(2002)\\
Rotational velocity of star V$_{\rm rot} \sin{i}$ & 3.75$\pm$1.25 \kms &
Queloz et al.~(2000)\\

\enddata
\end{deluxetable}


\clearpage
\begin{deluxetable}{lll}
\tabletypesize{\scriptsize} 
\tablecolumns{2}
\tablewidth{0pt}
\tablecaption{
HD~209458b Ephemeris \label{tbl-3}}
\tablehead{
\colhead{Parameter} & \colhead{Value \& Uncertainty}}

\startdata 
$T_0$  (HJD)  & 2452854.82545 $\pm \ 1.35 \times 10^{-4}$ \\
$Period$ (days) & 3.52474554    $\pm \ 1.8 \times 10^{-7}$ \\
\enddata
\end{deluxetable}

\begin{deluxetable}{llll}
\tabletypesize{\scriptsize}
\tablecolumns{4}
\tablewidth{0pt}
\tablecaption{HD~209458 System Parameters \label{tbl-4}}
\tablehead{
\colhead{Parameter} & \colhead{Estimate} & \colhead{} & \colhead{} }
\startdata
$M_{\rm star}$ (\Msun)           & 0.93 & 1.09 & 1.19\\
\hline
$R_{\rm star}$ (\Rsun)           & 1.085 & 1.144 & 1.178\\
$M_{\rm planet}$ (M$_{\rm Jup}$) & 0.593 & 0.657 & 0.697\\
$R_{\rm planet}$ (R$_{\rm Jup}$) & 1.279 & 1.349 & 1.388\\
Inclination (degrees)   & 86.668 & 86.668 & 86.668\\
$K_{\rm star}$ (m/s)    & 83.0 & 82.7 & 82.7\\
Orbital separation (AU) & 0.044 & 0.047 & 0.048\\
\hline
Limb-darkening coefficients & \\
$x (B)$ & 0.877 & 0.874 & 0.872 \\
$y (B)$ & 0.285 & 0.278 & 0.276 \\
$x (V)$ & 0.724 & 0.724 & 0.724 \\
$y (V)$ & 0.327 & 0.328 & 0.328 \\
$x (R)$ & 0.774 & 0.775 & 0.775 \\
$y (R)$ & 0.434 & 0.435 & 0.436 \\
$x (I)$ & 0.819 & 0.818 & 0.817 \\
$y (I)$ & 0.637 & 0.634 & 0.632 \\
\hline
$\chi^{2}$ (3434 degrees of freedom) & 4682.3 & 4681.4 & 4681.4\\
$\chi^{2}_{\nu}$ & 1.364 & 1.363 & 1.363 \\

\enddata
\end{deluxetable}


\clearpage
\begin{figure}
\plottwo{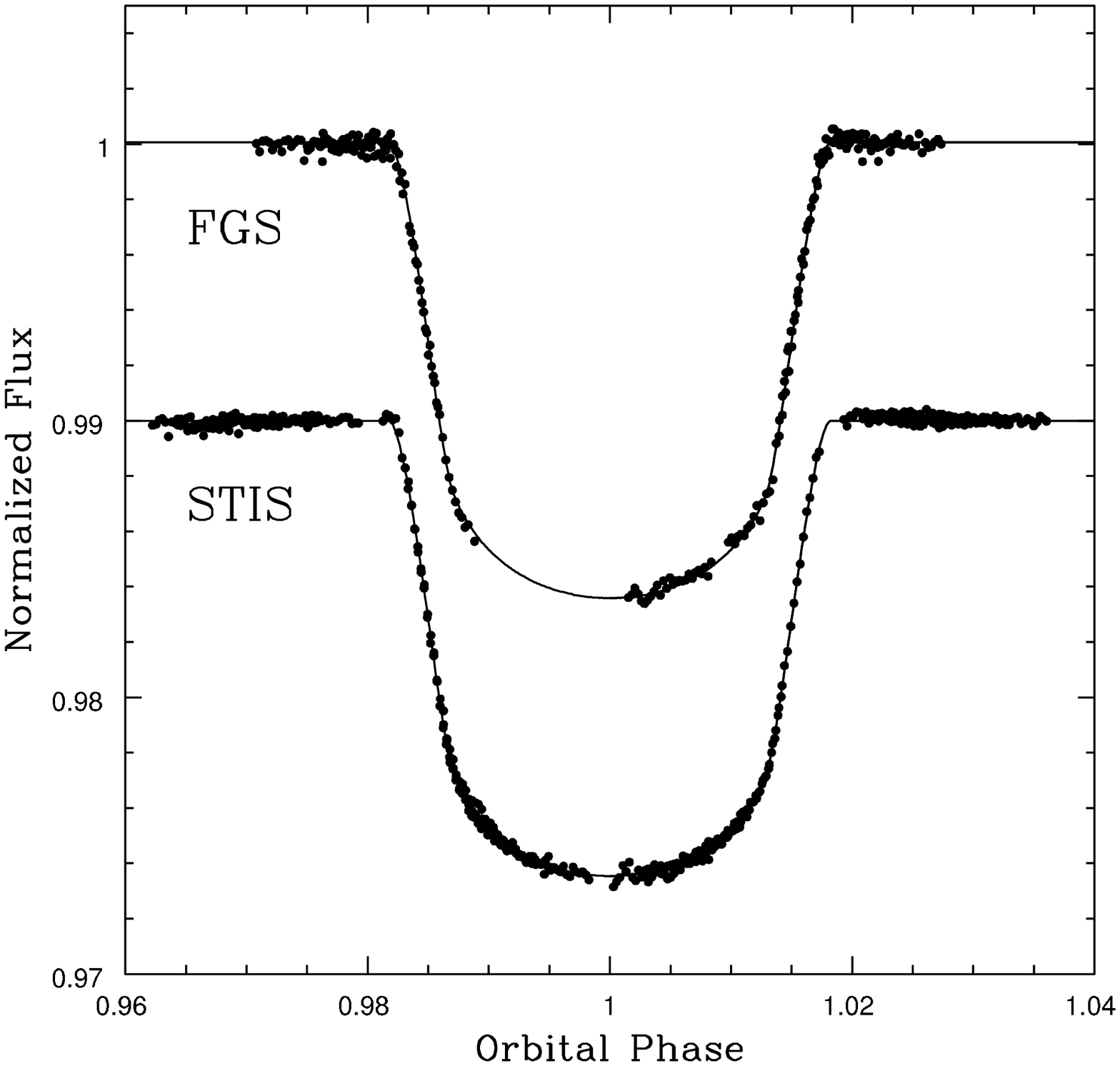}{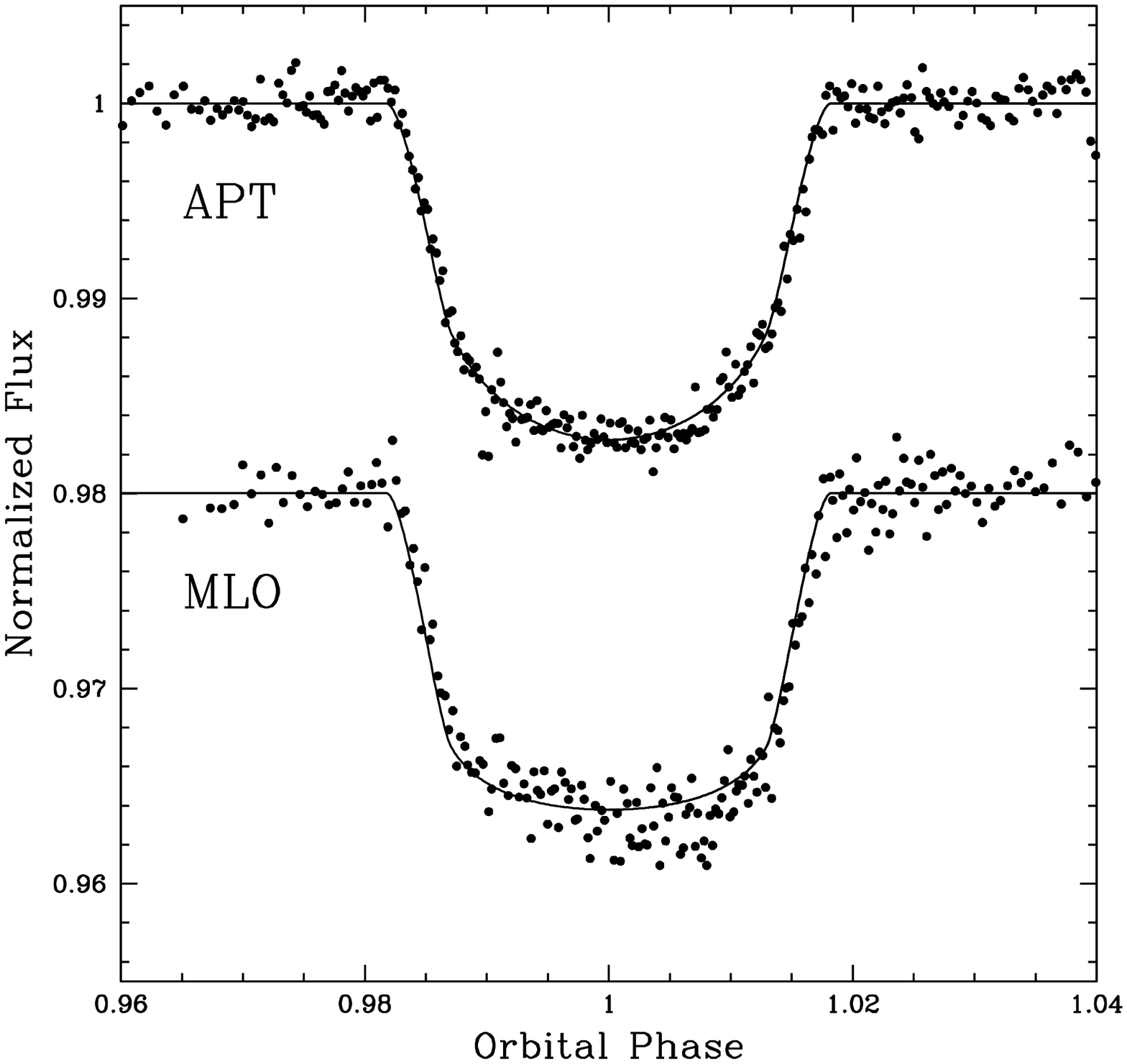}
\caption{Light curves and model fit, in normalized flux units (with
arbitrary offset) folded on our ephemeris.  The orbital phases shown
here span 6.7 hours and each tick mark corresponds to 25 minutes.  Left 
panel: HST FGS and STIS observations.  Right panel: Ground-based APT
and MLO observations. }
\end{figure}

\clearpage
\begin{figure}
\plottwo{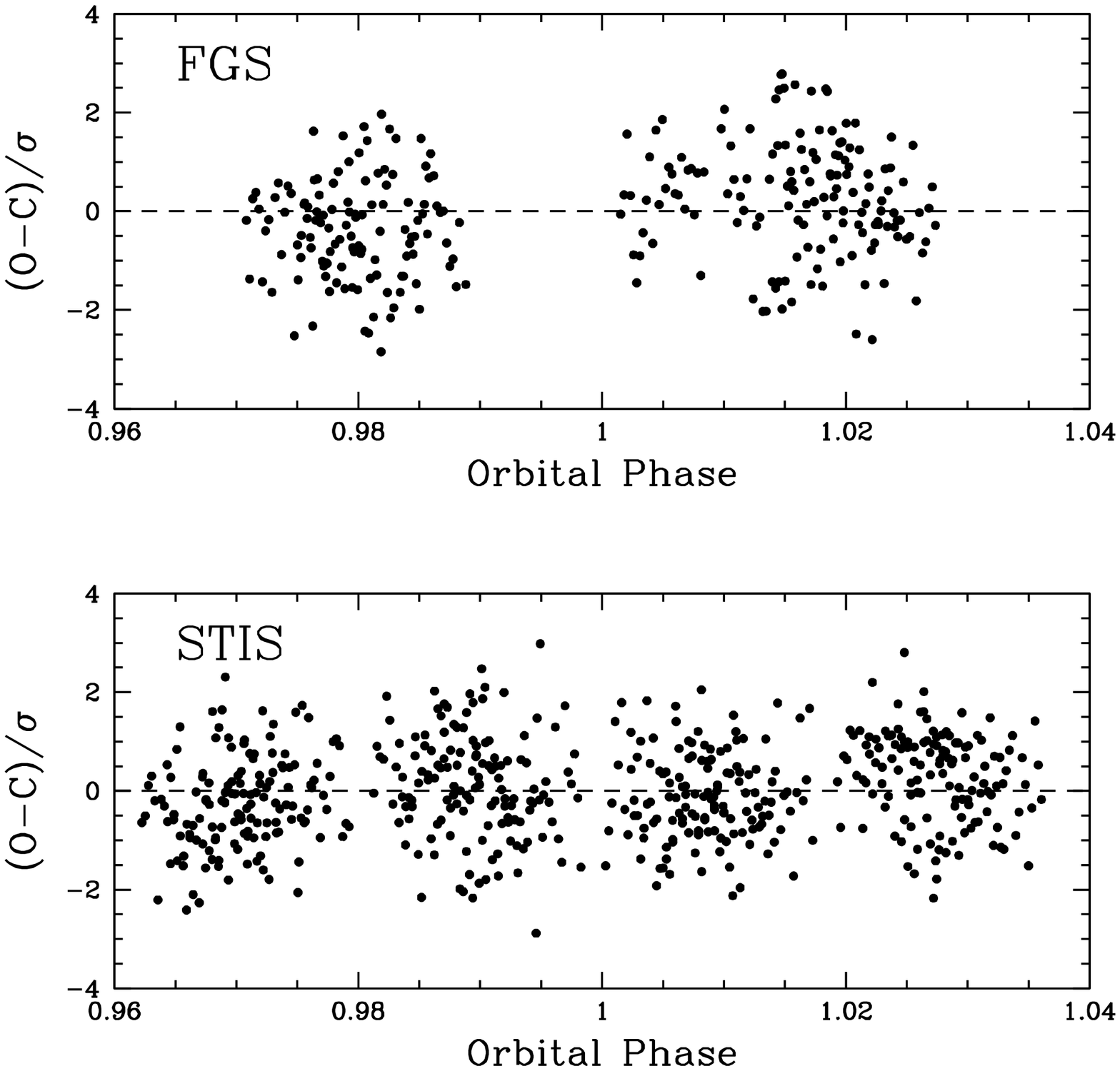}{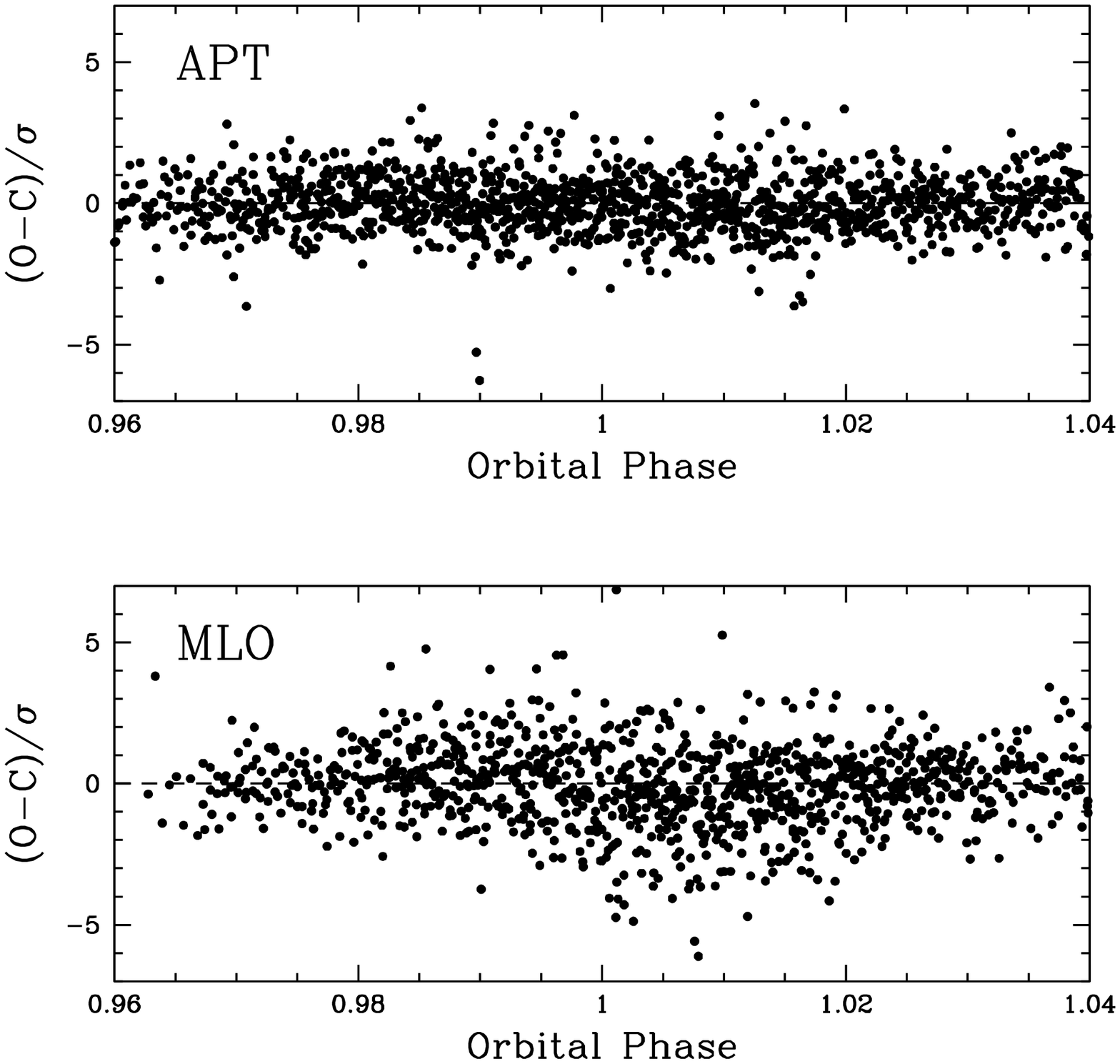}
\caption{Left panel: Normalized residuals ($\chi\equiv (O-C)/\sigma$) of
ELC model fit to FGS and STIS data sets.  Right panel: Normalized
residuals for APT and MLO data sets. }
\end{figure}

\clearpage
\begin{figure}
\plottwo{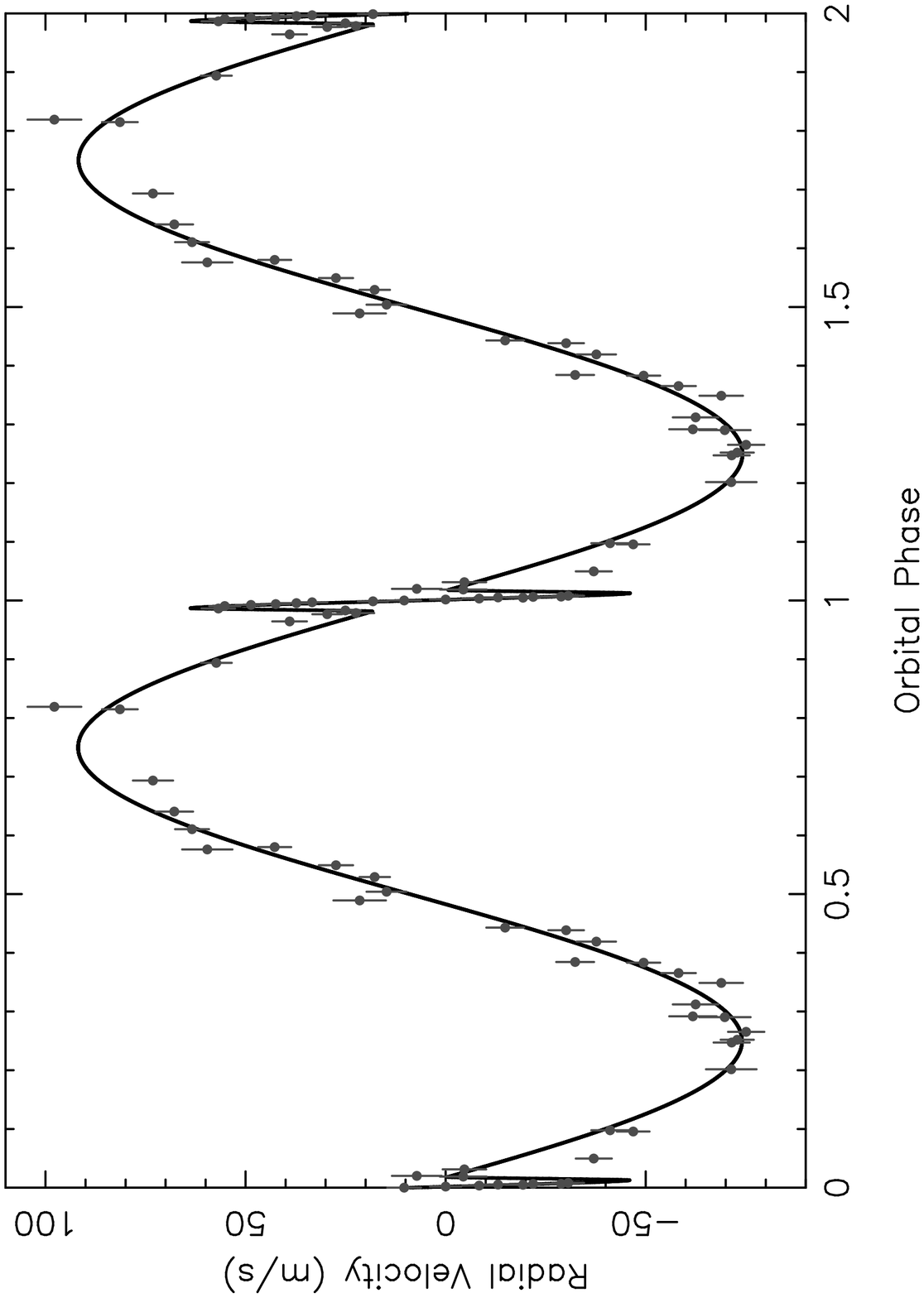}{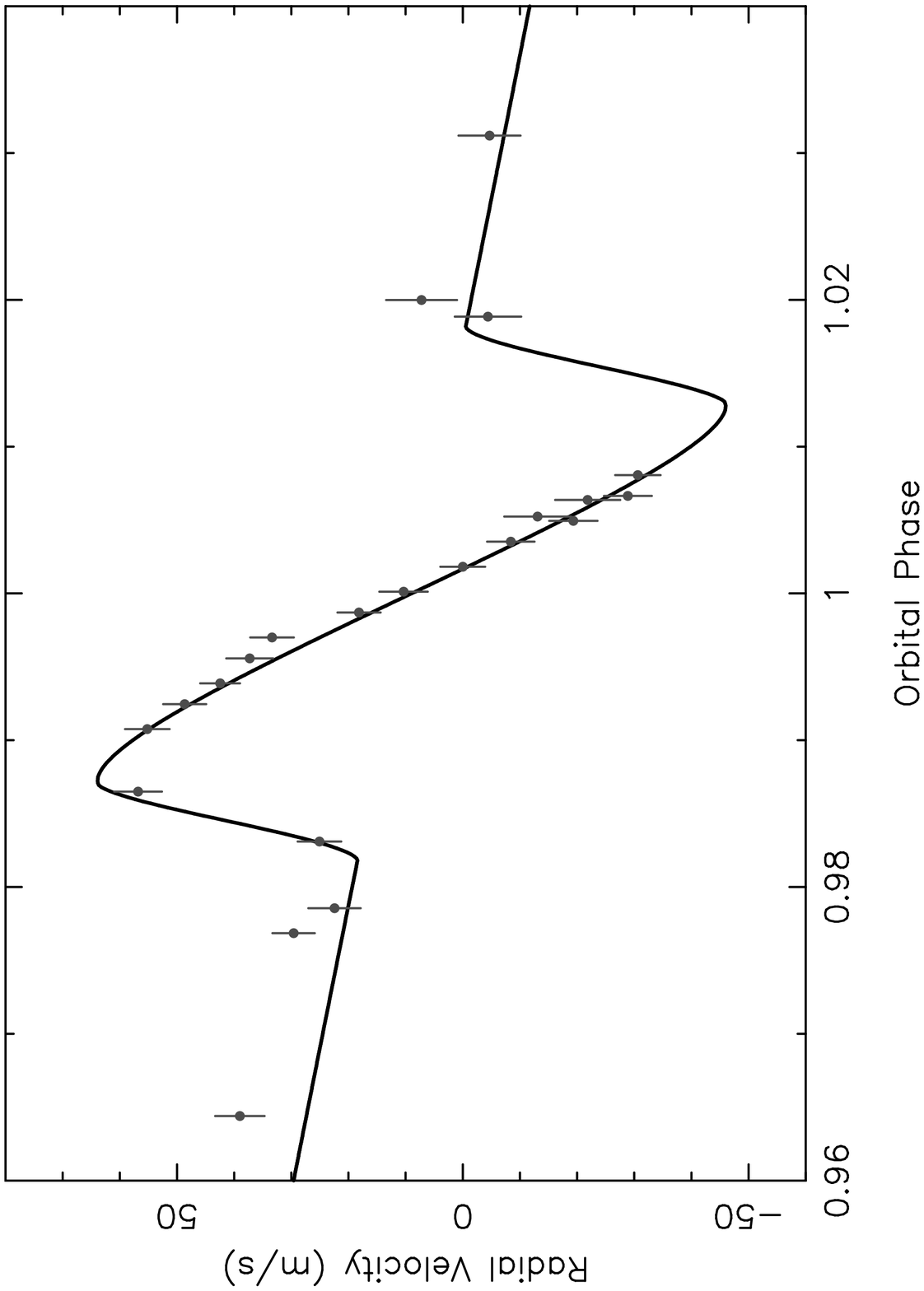}
\caption{Top panel: Radial velocity curve and model fit. Bottom panel:
Expanded view of the Rossiter effect during transit. 
[Editors: Please rotate figures clockwise 90 degrees.]
}
\end{figure}

\clearpage
\begin{figure}
\plotone{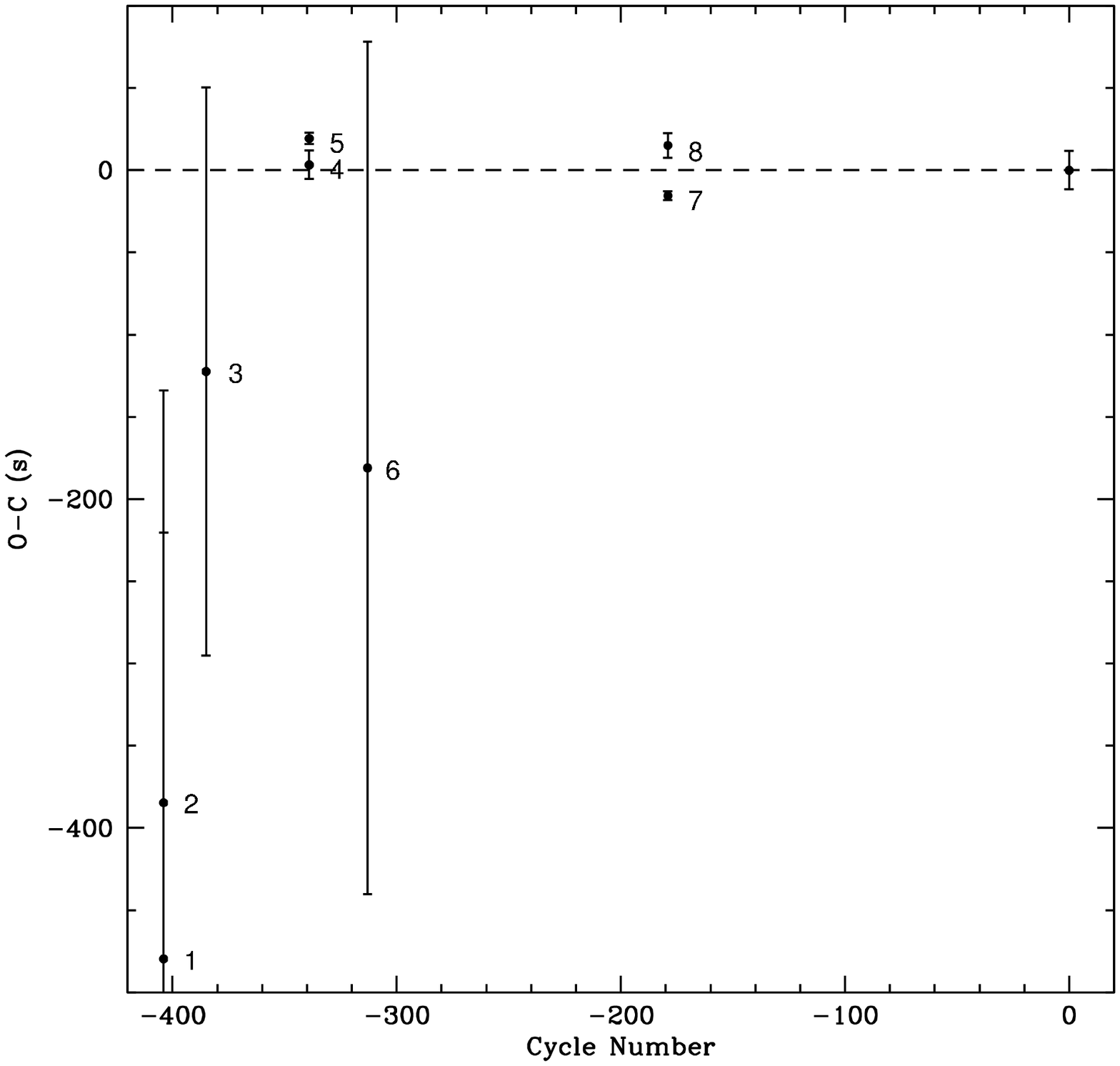}
\caption{
O-C diagram of published transit times. 1: \citet{charb00}, 
2: \citet{mazeh00}, 3: \citet{jha00}, 4: STIS \citep{brown01}, 
5: STIS \citep{schultz03}, 6: estimated from \citet{deeg01} Figure 1, 
7: FGS \& STIS \citep{schultz04}, 8: FGS \citep{schultz03}. 
Our result is shown at cycle 0.
}
\end{figure}

\clearpage
\begin{figure}
\epsscale{0.90}
\plotone{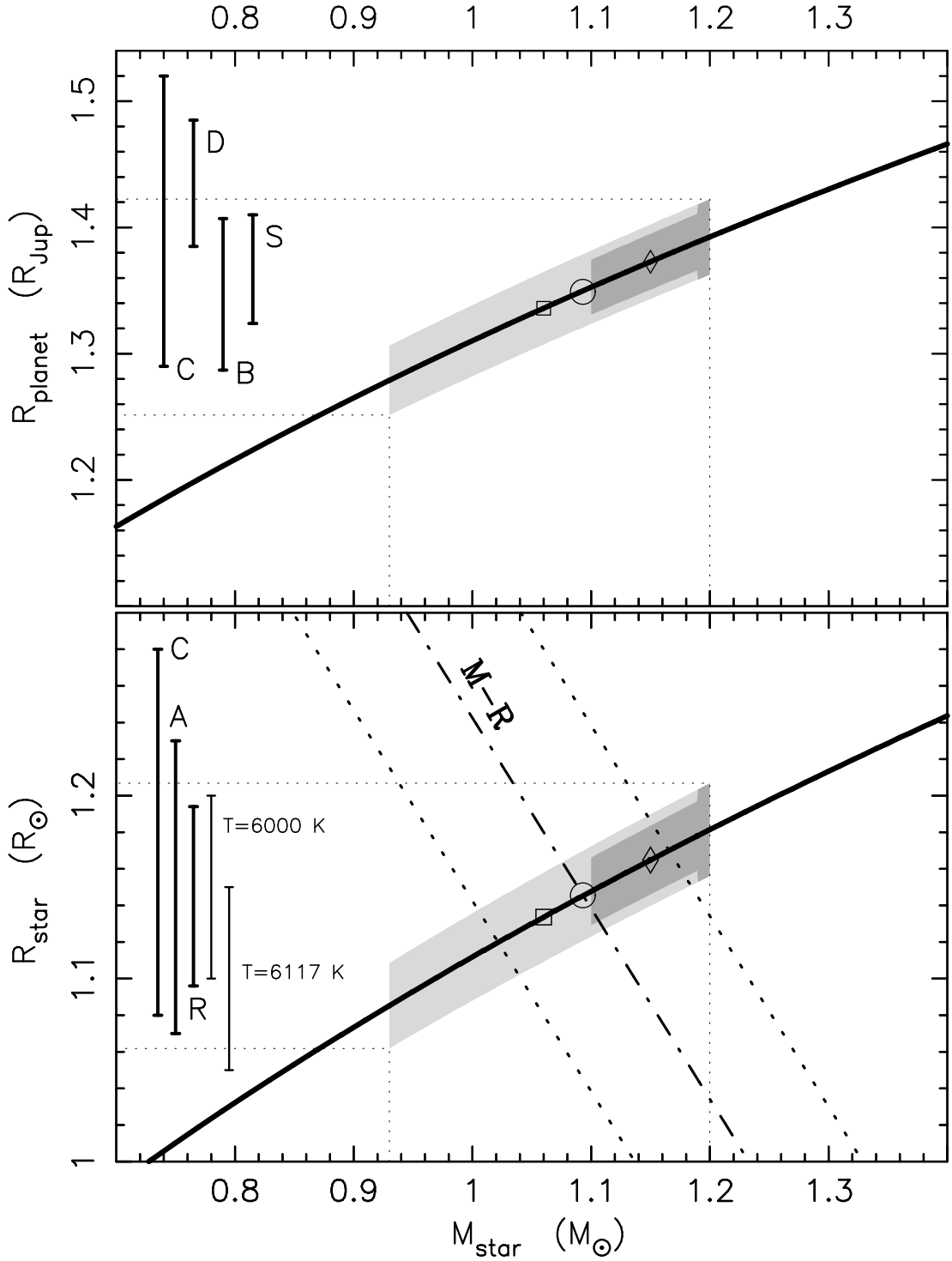}
\caption{
{\small{
Planetary and stellar radii versus stellar mass, as determined by the 
transit and radial velocity data. The shaded regions show acceptable 
solutions when a stellar mass constraint is used 
(light gray = \citet{cody02}, dark gray = \citet{santos04}).
For comparison, several published radius estimates are shown.
The dot--dash line in the lower panel is the mass--radius relationship 
of \citet{cody02}.
Being nearly orthogonal to the transit--derived relationship, this
strongly constrains the solutions.
}} 
}
\end{figure}


\clearpage
\begin{deluxetable}{lcc}
\tabletypesize{\scriptsize}
\tablecolumns{3}
\tablewidth{0pt}
\tablecaption{Appendix A. HD~209458 Radial Velocities \label{Appendix A}}
\tablehead{
\colhead{JD-2440000} & \colhead{Velocity (m/s)} & \colhead{Uncertainty 
(m/s)}}
\startdata
        11341.120 &     27.4 &     4.3 \\
        11368.941 &    -14.9 &     4.7 \\
        11372.134 &    -68.9 &     5.4 \\
        11373.056 &     63.4 &     4.2 \\
        11374.055 &     57.3 &     3.8 \\
        11410.012 &    -46.9 &     4.1 \\
        11410.963 &    -58.2 &     4.3 \\
        11411.933 &     67.8 &     4.7 \\
        11438.808 &    -75.1 &     4.5 \\
        11543.689 &    -4.4 &     5.8 \\
        11543.693 &     7.2 &     6.2 \\
        11550.691 &    -13.1 &     5.8 \\
        11550.695 &    -21.9 &     5.7 \\
        11551.696 &    -69.8 &     6.4 \\
        11551.701 &    -61.8 &     5.9 \\
        11552.703 &     59.6 &     6.3 \\
        11679.107 &    -30.1 &     4.4 \\
        11703.121 &    -72.9 &     4.1 \\
        11704.098 &     17.7 &     3.8 \\
        11705.105 &     81.4 &     4.4 \\
        11706.102 &    -41.1 &     4.7 \\
        11707.108 &    -49.5 &     4.1 \\
        11754.975 &     39.0 &     4.3 \\
        11755.019 &     29.6 &     3.7 \\
        11755.025 &     22.4 &     4.6 \\
        11755.041 &     25.1 &     3.8 \\
        11755.053 &     56.8 &     4.2 \\
        11755.068 &     55.2 &     3.9 \\
        11755.074 &     48.7 &     3.8 \\
        11755.079 &     42.5 &     3.5 \\
        11755.085 &     37.3 &     4.1 \\
        11755.090 &     33.4 &     3.8 \\
        11755.096 &     18.2 &     3.8 \\
        11755.101 &     10.4 &     4.2 \\
        11755.107 &     0.0 &     3.9 \\
        11755.113 &    -8.4 &     4.2 \\
        11755.118 &    -19.3 &     4.2 \\
        11755.124 &    -28.9 &     4.2 \\
        11755.129 &    -30.6 &     4.0 \\
        11755.972 &    -71.5 &     4.5 \\
        11792.791 &     73.2 &     5.0 \\
        11882.707 &    -71.4 &     6.3 \\
        11883.720 &     21.5 &     6.6 \\
        11899.731 &    -4.7 &     5.4 \\
        11900.721 &    -62.5 &     5.7 \\
        12063.112 &    -32.4 &     4.7 \\
        12102.004 &    -37.7 &     4.9 \\
        12446.128 &    -37.0 &     4.5 \\
        12514.965 &     42.7 &     4.1 \\
        12535.845 &     14.8 &     5.0 \\
        12575.730 &     97.8 &     6.7 \\
\enddata
\end{deluxetable}

\end{document}